\documentclass[twocolumn,showpacs,preprintnumbers,amsmath,amssymb]{revtex4}
\usepackage{graphicx}% Include figure files
\usepackage{dcolumn}% Align table columns on decimal point
\usepackage{bm}% bold math
\usepackage{subfigure}%

\begin{document}

%\preprint{APS/123-QED}

\title{Continuous-Variable Measurement-Device-Independent Quantum Key Distribution}% Force line breaks with \\

\author{Zhengyu Li$^{1,2}$}
\author{Yi-Chen Zhang$^3$}
\author{Feihu Xu$^4$}
\author{Xiang Peng$^{1\dag}$}
\author{Hong Guo$^{1,2}$}
\thanks{Corresponding author: hongguo@pku.edu.cn. \\$^{\dag}$Corresponding author: xiangpeng@pku.edu.cn.}

\affiliation{$^1$State Key Laboratory of Advanced Optical Communication Systems and Networks, School of Electronics Engineering and Computer Science and Center for Quantum Information Technology, Peking University, Beijing 100871, China}
\affiliation{$^2$Center for Computational Science \& Engineering (CCSE), Peking University, Beijing 100871, China.}
\affiliation{$^3$State Key Laboratory of Information Photonics and Optical Communications, Beijing University of Posts and Telecommunications, Beijing 100876, China}
\affiliation{$^4$Centre for Quantum Information and Quantum Control, Department of Physics and Department of Electrical $\&$ Computer Engineering, University of Toronto, Toronto, Ontario, M5S 3G4, Canada}
\date{September 2013}

\date{\today}% It is always \today, today,
             % but any date may be explicitly specified

\begin{abstract}
%Continuous-variable quantum key distribution (CV QKD) protocols suffer from the attacks against practically imperfect detectors. It is significant to remove the loopholes introduced by detectors to enhance the security of a practical system. Inspired by the idea of discrete-variable measurement-device-independent quantum key distribution (MDI-QKD) protocols, here we propose a continuous-variable MDI QKD protocol, in which the detections can be conducted by an \emph{untrusted} third party. The security analysis is based on the fact that the entanglement-based scheme of this protocol is equivalent to the one-way CV QKD with coherent states and heterodyne detection. A lower bound of the secret key rate against collective attacks is given. This new protocol can defend all attacks against detectors.

We propose a continuous-variable measurement-device-independent quantum key distribution (CV-MDI QKD) protocol, in which detection is conducted by an \emph{untrusted} third party. Our protocol can defend all detector side channels, which seriously threaten the security of a practical CV QKD system. Its security analysis against arbitrary collective attacks is derived based on the fact that the entanglement-based scheme of CV-MDI QKD is equivalent to the conventional CV QKD with coherent states and heterodyne detection. We find that the maximal total transmission distance is achieved by setting the \emph{untrusted} third party close to one of the legitimate users. Furthermore, an alternate detection scheme, a special application of CV-MDI QKD, is proposed to enhance the security of the standard CV QKD system.

\end{abstract}

\pacs{03.67.Dd, 03.67.Hk}% PACS, the Physics and Astronomy
\maketitle

\section{Introduction}
Quantum key distribution (QKD) \cite{Gisin_RevModPhys_2002, Scarani_RevModPhys_2009} can establish a secure key between two legitimate partners (Alice and Bob) through insecure quantum and classical channels. In recent last decades research on QKD has evolved rapidly. Some commercial systems are available in the market now \cite{Commerical}.
QKD has two main approaches: one is discrete-variable (DV) QKD, and the alternative is continuous-variable (CV) QKD \cite{Peter_RevModPhys.77.513_2005_CV, Weedbrook_RevModPhys.84.621_2012, Christian_PRA.87.022308}. CV QKD has the advantage of being compatible with standard telecommunication technology, especially no request on single photon detectors. A Gaussian modulated CV QKD protocol using coherent states \cite{Grosshans_PhysRevLett_2002_GG02,Weedbrook_NoSwitch_PhysRevLett_2004} has been proved to be secure against arbitrary collective attacks \cite{Grosshans_PhysRevLett.94.020504_2005_CollAtt, Acin_PhysRevLett.94.020505_2005_CollAtt}, which is optimal in both the asymptotic case \cite{Renner_PhysRevLett_2009_CVdeFinetti} and the finite size regime \cite{PhysRevLett.109.100502_2012_FinitSize, Leverrier_PhysRevLett_2013_generalAtt}.  A recent experiment has successfully distributed secure keys over an 80-km optical fiber \cite{Jouguet_NPhoton_2013}, showing the potential of long distance communication using CV QKD protocols.

Generally speaking, the theoretical security analysis of QKD relies on some ideal theoretical models. However, the practical devices often have some imperfections and deviate from the theoretical models. Thus the mismatch between practical devices and their idealized models may open security loopholes, which make the practical systems vulnerable to attacks \cite{Hoi_Springer_2009}. In DV QKD systems, various types of attacks against imperfect devices were proposed, among which the attacks against the single photon detector are the most significant ones \cite{HoiKwong_PhysRevA.78.042333_2008_hacking, Lydersen_NaturePhotonics_2010, Makarov_NatureCommu.2.349_2011}. Recently in CV QKD systems, several attack strategies against practical detectors were also proposed \cite{Lutkenhaus_PhysRevA.77.032303_LO, Jouguetn_PhysRevA_2013,MaXiangchun_Phys.Rev.A_2013_LOfluc, Han_PhysRevA_2013, PhysRevA.87.2013_wavelength, saturation_attack_2013}. For example, the calibration attack \cite{Jouguetn_PhysRevA_2013} and local oscillator (LO) fluctuation attack \cite{MaXiangchun_Phys.Rev.A_2013_LOfluc} take advantage of modifying LO to manipulate the measurement results, which will make Alice and Bob overestimate the secret key rate. The wavelength attack \cite{Han_PhysRevA_2013, PhysRevA.87.2013_wavelength} allows the eavesdropper to launch an intercept-resend attack because of the wavelength-dependent property of the fiber beamsplitter used in the heterodyne detector. The saturation attack \cite{saturation_attack_2013} can force Alice and Bob to underestimate the excess noise by saturating the homodyne detector, which can hide the presence of an intercept-resend attack.

A natural attempt to remove these attacks in a CV QKD system was to characterize the specific loophole and find a countermeasure. For instance, Jouguet \emph{et al.} proposed an efficient countermeasure against the calibration attack by monitoring the LO \cite{Jouguetn_PhysRevA_2013}. Once an attack is known, prevention is usually simple. However, it is difficult to fully characterize real detectors and account for all loopholes. Therefore figuring out how to defend against general attacks on detectors in practical systems becomes critical in CV QKD.

Inspired by the novel detector-attack-immune protocols, i.e. measurement-device-independent (MDI) QKD protocols \cite{LO_PhysRevLett_2012,Pirandola_PhysRevLett.108.130502_SiChFree, MaXF_PhysRevA.86.062319_2012_PhEncod}, which were well analyzed in theory \cite{ WangXB_PhysRevA.87.012320_decoy, PracMDI_Feihu_NJP.15.113007.2013, FiniteMDI_Curty_arXiv, WangXB_arXiv.1308.5677_decoy} and successfully demonstrated in experiments \cite{PhysRevLett.111.130501, Silva_PhysRevA_2013, Pan_PhysRevLett_2013, Feihu_arXiv_2013_exper},  here we propose a CV-MDI QKD protocol which can also defend all detector side channels. The main idea is that both Alice and Bob are senders and an \emph{untrusted} third party, named Charlie, is introduced to realize the measurement. Such measurement results will be used by Alice and Bob in the post-processing step to generate secure keys.

%In this paper, we propose a continuous-variable measurement-device-independent (CV-MDI) QKD protocol which can defend all attacks against detectors.  The main idea is that both Alice and Bob are senders and an \emph{untrusted} third party, named Charlie, is introduced to realize the measurement. Such measurement results will be used by Alice and Bob in post-processing step to generate secure keys. This idea has already been used in DV QKD for theoretical analysis\cite{LO_PhysRevLett_2012, Pirandola_PhysRevLett.108.130502_SiChFree, MaXF_PhysRevA.86.062319_2012_PhEncod,PhysRevLett.111.130501, Silva_PhysRevA_2013, Pan_PhysRevLett_2013, Feihu_arXiv_2013_exper}.

By introducing the equivalent entanglement-based (EB) scheme of this protocol, we show the security analysis against arbitrary collective attacks, which is based on the fact that the entanglement-based scheme of CV-MDI QKD is equivalent to the CV QKD with coherent states and heterodyne detection \cite{Weedbrook_NoSwitch_PhysRevLett_2004}. A corresponding prepare-and-measure (PM) scheme is proposed for implementation. Moreover, the performance of our protocol against collective entangling cloner attack is presented via numerical simulations. When the distance between Alice and Charlie equals that between Bob and Charlie (symmetric case), the transmission distance is below 10 km. However, in the asymmetric case the transmission distance can be improved, reaching 80 km under certain conditions. This demonstrates the feasibility of our scheme.

This paper is organized as follows: In Sec. II, the detailed descriptions of both the EB and PM schemes of CV-MDI QKD are given. In Sec. III, we present the security analysis for the CV-MDI QKD protocol. In Sec. IV, we show the numerical simulation results of the secret key rate, and discuss the performance and potential applications.

\section{Continuous-variable Measurement-device-independent QKD Protocol}

\begin{figure}[t]
\includegraphics[width=3.4in]{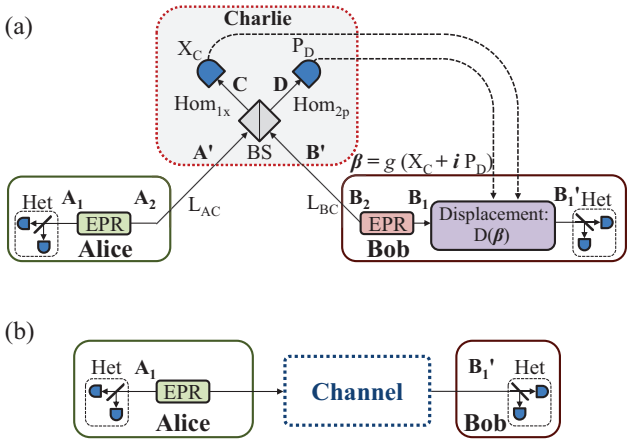}
\caption{(Color online)(a) EB scheme of the CV-MDI QKD protocol. EPR: two-mode squeezed state. Het: heterodyne detection. $\rm{Hom_{1x}}$: homodyne detection of measuring the \emph{x} quadrature. $\rm{Hom_{2p}}$: homodyne detection of measuring the \emph{p} quadrature. $X_C, P_D$: measurement results of $\rm{Hom_{1x}}$ and $\rm{Hom_{2p}}$, respectively. BS: 50:50 beamsplitter. (b) Equivalent one-way model of the EB scheme under the assumptions that Eve controls channels, Charlie and Bob's EPR state, and the displacement in (a). }\label{EB}
\end{figure}

In the PM scheme of a standard QKD protocol such as Bennett-Brassard 1984 (BB84) \cite{Bennett_IEEE_1984_BB84}, Alice randomly prepares an encoded quantum state and sends it to Bob for detection. This PM scheme can be formulated in an EB version of the protocol as follows: Alice first creates an entangled state such as a Bell state, and afterwards, she measures one mode of this entangled state in a certain basis, thereby producing the correct state for the other mode that is sent to Bob.

In practice, the PM scheme is usually easy to apply, while the equivalent EB scheme is convenient for security analysis.
The EB scheme of DV-MDI QKD can be seen as a one-way protocol using entanglement swapping as an untrusted quantum relay \cite{LO_PhysRevLett_2012, Pirandola_PhysRevLett.108.130502_SiChFree}. Here we use the same idea in our CV-MDI QKD protocol, which exploits the continuous variable entanglement swapping \cite{Ralph_PhysRevLett_1999, Furusawa_Science.282.706_1998_Telep}.
The EB scheme of the CV-MDI QKD protocol shown in Fig. \ref{EB}(a) is described as follows.

{\emph{Step 1.}} Alice generates one two-mode squeezed (TMS) state and keeps mode $A_1$ while sending the other mode, $A_2$, to an \emph{untrusted} third party (Charlie) through the channel with length $L_{AC}$. Bob generates another TMS state and keeps the mode $B_1$ while sending the other mode, $B_2$, to Charlie through another channel with length $L_{BC}$.

{\emph{Step 2.}} Modes $A'$ and $B'$ received by Charlie interfere at a beam splitter (BS) with two output modes $C$ and $D$. Then both the \emph{x} quadrature of $C$ and \emph{p} quadrature of $D$ are measured by homodyne detections, and the measurement results $\{X_C, P_D\}$ are publicly announced by Charlie.

\begin{figure}[t]
\includegraphics[height=2.5in]{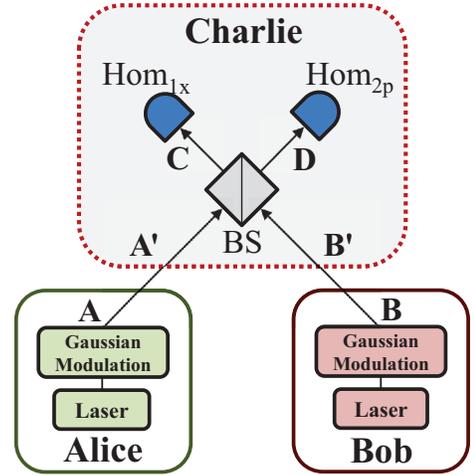}
\caption{(Color online) PM scheme of CV-MDI QKD. $\rm{Hom_{1x}}$: homodyne detection of measuring the \emph{x} quadrature. $\rm{Hom_{2p}}$: homodyne detection of measuring the \emph{p} quadrature. BS: 50:50 beam splitter.}\label{PM}
\end{figure}

{\emph{Step 3.}} After receiving Charlie's measurement results, Bob displaces mode $B_1$ by operation $\hat D\left( \beta  \right)$ and gets ${\hat\rho _{{B_1}'}} = \hat D\left( \beta  \right){\hat\rho _{{B_1}}}{{\hat D}^\dag }\left( \beta  \right)$, where $\hat\rho_{X}$ represents the density matrix of mode $X$, $\beta = {g\left( {{X_C} + i{P_D}} \right)} $ and $g$ represents the gain of the displacement. Then Bob measures mode $B_1'$ to get the final data $\left\{X_B,P_B\right\}$ using heterodyne detection. Alice measures mode $A_1$ to get the final data $\left\{X_A, P_A\right\}$ using heterodyne detection.

{\emph{Step 4.}} Alice and Bob use an authenticated public channel to finish
the parameter estimation, information reconciliation, and privacy amplification steps.

After Charlie's measurements and Bob's displacement, mode $A_1$ and mode $B_1'$ become entangled \cite{Ralph_PhysRevLett_1999}. Therefore, after both Alice's and Bob's heterodyne detections, their final data are correlated. The equivalent PM scheme is shown in Fig. \ref{PM}, which is described as follows.

{\emph{Step 1.}} Alice randomly prepares a coherent state $\left| {{x_A} + i{p_A}} \right\rangle $, where $x_A$ and $p_A$ are Gaussian distributed with variance $V_A-1$. Bob randomly prepares another coherent state $\left| {{x_B} + i{p_B}} \right\rangle $, where $x_B$ and $p_B$ are Gaussian distributed with variance $V_B-1$. Both Alice and Bob send their coherent states to Charlie through two different channels.

{\emph{Step 2.}} The two modes ($A'$ and $B'$) received by Charlie interfere at a BS with two output modes $C$ and $D$. Then both the \emph{x} quadrature of $C$ and \emph{p} quadrature of $D$ are measured by homodyne detections, and the measurement results $\{X_C, P_D\}$ are publicly announced by Charlie.

{\emph{Step 3.}} When Alice and Bob receive Charlie's measurement results, Bob modifies his data as ${X_B} = {x_B} + {k}{{X}_C},{P_B} = {p_B} - {k}{{P}_D}$, while Alice keeps hers unchanged, ${X_A} = {x_A},{P_A} = {p_A}$. $k$ is the amplification coefficient related to channel loss (the relationship between $k$ and $g$ in the EB scheme is shown in Appendix A).

{\emph{Step 4.}} Alice and Bob use an authenticated public channel to finish the parameter estimation, information reconciliation, and privacy amplification steps.

In the PM scheme, Alice and Bob prepare coherent states independently and do not require any measurements inside them. The third party, Charlie, is totally \emph{untrusted}. That is why the protocol can be called measurement device independent.
See Appendix A for the proof of equivalence between the EB and PM schemes.

\section{Security Analysis}
It is well known that the security of a PM scheme is equivalent to that of the corresponding EB scheme for a QKD protocol \cite{Grosshans_QIC_2003_entanglingCloner}.
In the EB scheme in Fig. \ref{EB}(a), if one further assumes that both Bob's initial TMS state and the displacement operation inside himself are also \emph{untrusted}, 
then the protocol could be seen as the well-known one-way CV QKD protocol using coherent states and heterodyne detection \cite{Weedbrook_NoSwitch_PhysRevLett_2004}. 
The equivalent one-way model is shown in Fig. \ref{EB}(b). Thus the EB scheme of CV-MDI QKD is just one specific case of the equivalent one-way model with more constraints on Eve. Therefore, the secret key rate of the equivalent one-way model should be a lower bound of the EB scheme. Suppose the secret key rates
(with reverse reconciliation) of the EB scheme of CV-MDI QKD and the equivalent one-way model are, respectively, $K_1^R$ and $K_2^R$; then $K_2^R \le K_1^R$.
Although $K_2^R$ is not very tight, it is easy to calculate \cite{Winter_Proc.R.Soc.A.461.207_2005_SecKeyRate, Patron_thesis_2007} %SecretKeyRate
$$ {K_2^{R}} = \beta_R I\left( {{X_A, P_A}:{X_B, P_B}} \right) - \chi_{2} \left( {X_B, P_B}:E \right),$$ where $I\left( {{X_A, P_A}:{X_B, P_B}} \right)$ is the classical mutual information between Alice and Bob, $\beta_R$ is the reconciliation efficiency, and $\chi_{2} \left( {{X_B, P_B}:E} \right)$ is the Holevo bound \cite{Neilsen_2000} of the mutual information between Bob and Eve. Also $\chi_2 \left( {{X_B, P_B}:E} \right) = S\left( \hat \rho_E\right) - S\left(\hat \rho_E|X_B, P_B\right)$, where $S(\hat\rho_E)$ is the von Neumann entropy of the quantum state $\hat\rho_E$.  $S\left( {{{\hat \rho }_E}|{X_B},{P_B}} \right) = \int {P\left( {{X_B},{P_B}} \right)S\left( {\hat \rho _{E|{{X_B},{P_B}}}} \right)d{X_B}d{P_B}}$, ${P\left( {{X_B},{P_B}}\right)}$ is the probability of getting the measurement results ${X_B},{P_B}$ and $\hat \rho _{E|{{X_B},{P_B}}}$ is the conditional density matrix.

Generally, one can assume that Eve can purify the whole system, which means $\chi_2 \left( {{X_B, P_B}:E} \right)=\chi_2 \left( {X_B, P_B}:{{A_1}{B_1'}} \right)$ \cite{Grosshans_PhysRevLett.94.020504_2005_CollAtt, Acin_PhysRevLett.94.020505_2005_CollAtt}. Thus, the secret key rate can be written as
\begin{eqnarray}
{K_2^{R}} &=& \beta_RI\left( {{X_A, P_A}:{X_B, P_B}} \right)  \nonumber \\
&\ & - \left[S\left( \hat \rho_{{A_1}{B_1'}} \right) - S\left( \hat \rho_{{A_1}{B_1'}}|{X_B, P_B} \right)\right].
\end{eqnarray}

Based on the theorem of the optimality of Gaussian collective attacks \cite{Acin_PhysRevLett.97.190502_2006_GAO,Patron_OGA_PhysRevLett_2006}, the upper bound of $\chi_2 \left( {X_B, P_B}:{{A_1}{B_1'}} \right)$ is only determined by the covariance matrix ${\gamma_{{A_1}{B_1'}}}$ of the quantum state ${\hat\rho _{{A_1}{B_1'}}}$.
In a practical experiment, ${\gamma _{{A_1}{B_1'}}}$ can be calculated through the parameter estimation step.

From the analysis above, the secret key rate $K_2^R$ is based on the assumption that Eve controls Charlie, and it's actually calculable in a practical experiment. Therefore, the CV-MDI QKD protocol using $K_2^R$ as the secret key rate is immune to all collective attacks against detectors.

\begin{figure}[t]
\includegraphics[width=3.5in]{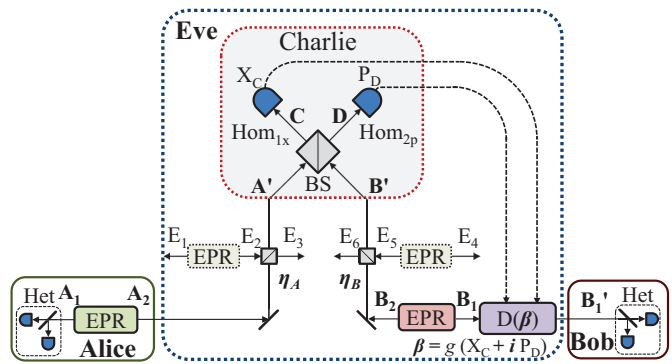}
\caption{(Color online) EB scheme of CV-MDI QKD with two independent entangled-cloner attacks. After two channels, mode $A_2$ becomes $A'$, and mode $B_2$ becomes $B'$. Charlie keeps the same structure as in Fig. \ref{EB}.}\label{EBcal}
\end{figure}

\section{Simulation results and Discussion}
\subsection{Numerical simulation}
As discussed above, in experiment, Alice and Bob can get the covariance matrix ${\gamma _{{A_1}{B_1'}}}$ through the parameter estimation step.
Then they can calculate the secret key rate $K_2^R$. In numerical simulation,
a model to simulate the CV-MDI QKD protocol is provided, including what the channels are and what Charlie does. The model is shown in Fig. \ref{EBcal}.

We assume the channels from Alice to Charlie and Bob to Charlie are under two independent entangling cloner \cite{Grosshans_QIC_2003_entanglingCloner} attacks, and Charlie preforms a standard CV entanglement-swapping process as the EB scheme requires. All the simulations in this paper are under this model.

We should point out that Eve's attack described here is not the optimal one.
\begin{figure}[t]
\includegraphics[width=3.3in]{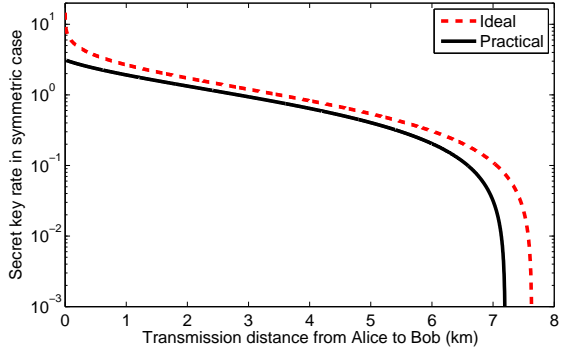}
\caption{(Color online) Secret key rate in the symmetric case where ${L_{AC} = L_{BC}}$. The red dashed line is under the ideal condition that variance $V_A = V_B = 10^5$ and zero excess noise. The black solid line is under the practical condition that variance $V_A = V_B = 40$ and excess noises $\varepsilon_A = \varepsilon_B = 0.002$ \cite{Jouguet_NPhoton_2013}. Both reconciliation efficiencies are $\beta_R = 1.$ }\label{sim_sym}
\end{figure}
The entangling cloner attack is usually used to model a Gaussian channel affected by the environment (Eve) and is analyzed to get a sense of a protocol's performance in experiment. In Fig. \ref{EBcal}, the channel parameters transmittance and excess noise on Alice's side (Bob's side) are $\eta_A$ ($\eta_B$) and $\varepsilon_A$ ($\varepsilon_B$). We assume that both channel losses are $\alpha = 0.2$~dB/km, and thus $\eta_A = 10^{-\alpha L_{AC}/10}$ and $\eta_B = 10^{-\alpha L_{BC}/10}$. The quadratures' relations are shown in Appendix B.
The covariance matrix $\gamma _{{A_1}{B_1'}}$ is
\begin{equation}
\left( {\begin{array}{*{20}{c}}
 {{V_A}{{\emph{I}}_2}}&{\sqrt {T\left( {V_A^2 - 1} \right)} {\sigma _z}}\\
{\sqrt {T\left( {V_A^2 - 1} \right)} \cdot{\sigma _z}}&{\left[ {T\left( {{V_A} - 1} \right) + 1 + T\varepsilon '} \right]{{\emph{I}}_2}}
\end{array}} \right)\label{cm}
\end{equation}
where
\begin{equation}
{{\emph{I}}_2} = \left( {\begin{array}{*{20}{c}}
1&0\\
0&1
\end{array}} \right){\rm{,\ \ \ }}{\sigma _z} = \left( {\begin{array}{*{20}{c}}
1&0\\
0&{ - 1}
\end{array}} \right),T = \frac{{{\eta _A}}}{2}{g^2},\label{T}
\end{equation}
 and
\begin{eqnarray}
\varepsilon ' = 1 &+& \frac{1}{{{\eta _A}}}\left[ {{\eta _B}\left( {{\chi _B} - 1} \right) + {\eta _A}{\chi _A}} \right] \nonumber \\
 &+& \frac{1}{{{\eta _A}}}{\left( {\frac{{\sqrt 2 }}{g}\sqrt {{V_B} - 1}  - \sqrt {{\eta _B}} \sqrt {{V_B} + 1} } \right)^2}.\label{Epsilon}
\end{eqnarray}

\noindent $\chi_A = \frac{1-\eta_A}{\eta_A} + \varepsilon_A$, $\chi_B = \frac{1-\eta_B}{\eta_B} + \varepsilon_B$, and \emph{g} is the gain of displacement.

Comparing the covariance matrix (\ref{cm}) with the one-way protocol, we can find that ${\varepsilon}'$ represents the equivalent excess noise of the equivalent one-way model of CV-MDI QKD. Here we choose $g = \sqrt {\frac{2}{{{\eta _B}}}} \sqrt {\frac{{{V_B} - 1}}{{{V_B} + 1}}}$ to minimize the equivalent excess noise; thus we have
\begin{equation}
\varepsilon ' = {\varepsilon _A} + \frac{1}{{{\eta _A}}}\left[ {{\eta _B}\left( {{\varepsilon _B} - 2} \right) + 2} \right].\label{EquEN}
\end{equation}

\begin{figure}[t]
%\subfigure[\ ]{\includegraphics[width=3.0in]{fig6_displace_one-way_unbalance3.eps}}
\subfigure[\ ]{\includegraphics[width=3.3in]{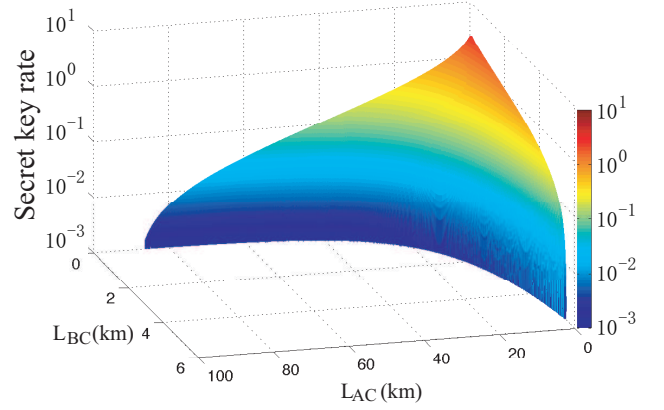}}
\subfigure[\ ]{\includegraphics[width=3.3in]{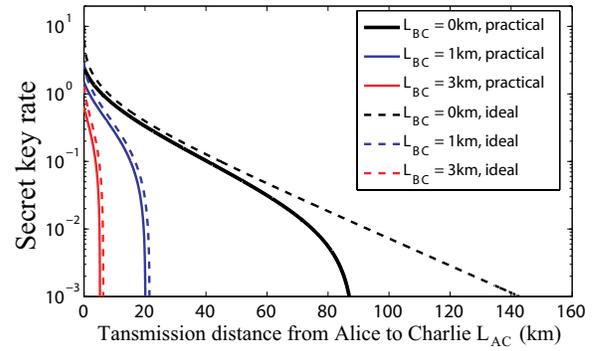}}
\caption{(Color online) (a) Secret key rate vs transmission distance from Alice to Charlie $L_{AC}$ and from Bob to Charlie $L_{BC}$, under the practical conditions that $V_A = V_B = 40$ and $\varepsilon_A = \varepsilon_B = 0.002$ \cite{Jouguet_NPhoton_2013}. (b) Secret key rate vs $L_{AC}$ given different $L_{BC}$. The dashed lines are under the ideal conditions that $V_A = V_B = 10^5$ and $\varepsilon_A = \varepsilon_B = 0$. The solid lines are under the same practical conditions as in (a).
$L_{BC}$ is $0$~km for the top black solid and dashed lines \cite{ZeroTransDis}, $1$~km for the middle blue solid and dashed lines, and $3$~km for the bottom red solid and dashed lines. Reconciliation efficiency is set to $\beta_R = 1$ for all simulations.
}\label{sim2}
\end{figure}

We first consider the perfect reconciliation efficiency case $\beta_R = 1$. The simulation result in Fig. \ref{sim_sym} is the secret key rate of the symmetric case, which means $L_{AC}=L_{BC}$.
 The red dashed line is under the conditions that variance $V_A = V_B = 10^5$ and excess noise $\varepsilon_A = \varepsilon_B = 0$, and the black solid line is under the conditions that variance $V_A = V_B = 40$ and excess noises $\varepsilon_A = \varepsilon_B = 0.002$, which is reasonable according to experiment \cite{Jouguet_NPhoton_2013}.

From Fig. 4, we can see that in both cases, the maximal total transmission distance ($L_{AB} = L_{AC}+L_{BC}$) is around 7km, referring to a 1.4 dB loss. If $L_{AC} = 3.5$km, even if there is no excess noise in either channel, the equivalent excess noise in Eq. (\ref{EquEN}) is still around 0.35. This is a very large value, and it will increase quickly if the transmission distance gets longer. That's why we cannot extract the secret key at a longer distance for the symmetric case. Hence the symmetric case can be useful in short-distance communications. In the next section, we will discuss how to extend the transmission distance.

\subsection{Discussion and application}
Equation (\ref{EquEN}) indicates that $\varepsilon'$ is not symmetric if permuting $\eta_A$ and $\eta_B$ because the postprocessing step is not symmetric because only Bob modifies his data while Alice keeps hers unchanged. This means that the symmetric case cannot result in an optimal system performance.

\begin{figure}[t]
\subfigure[\ ]{\includegraphics[width=2.8in]{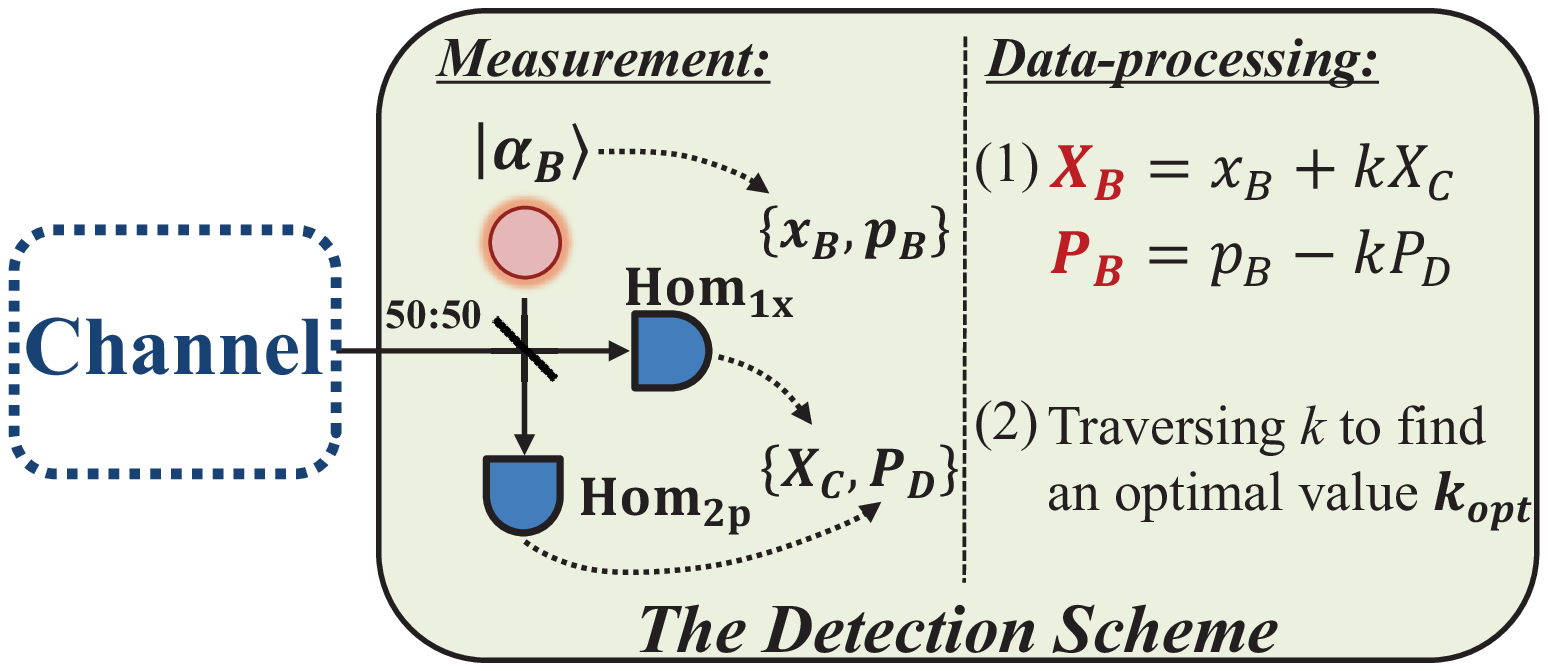}}
\subfigure[\ ]{\includegraphics[width=3.3in]{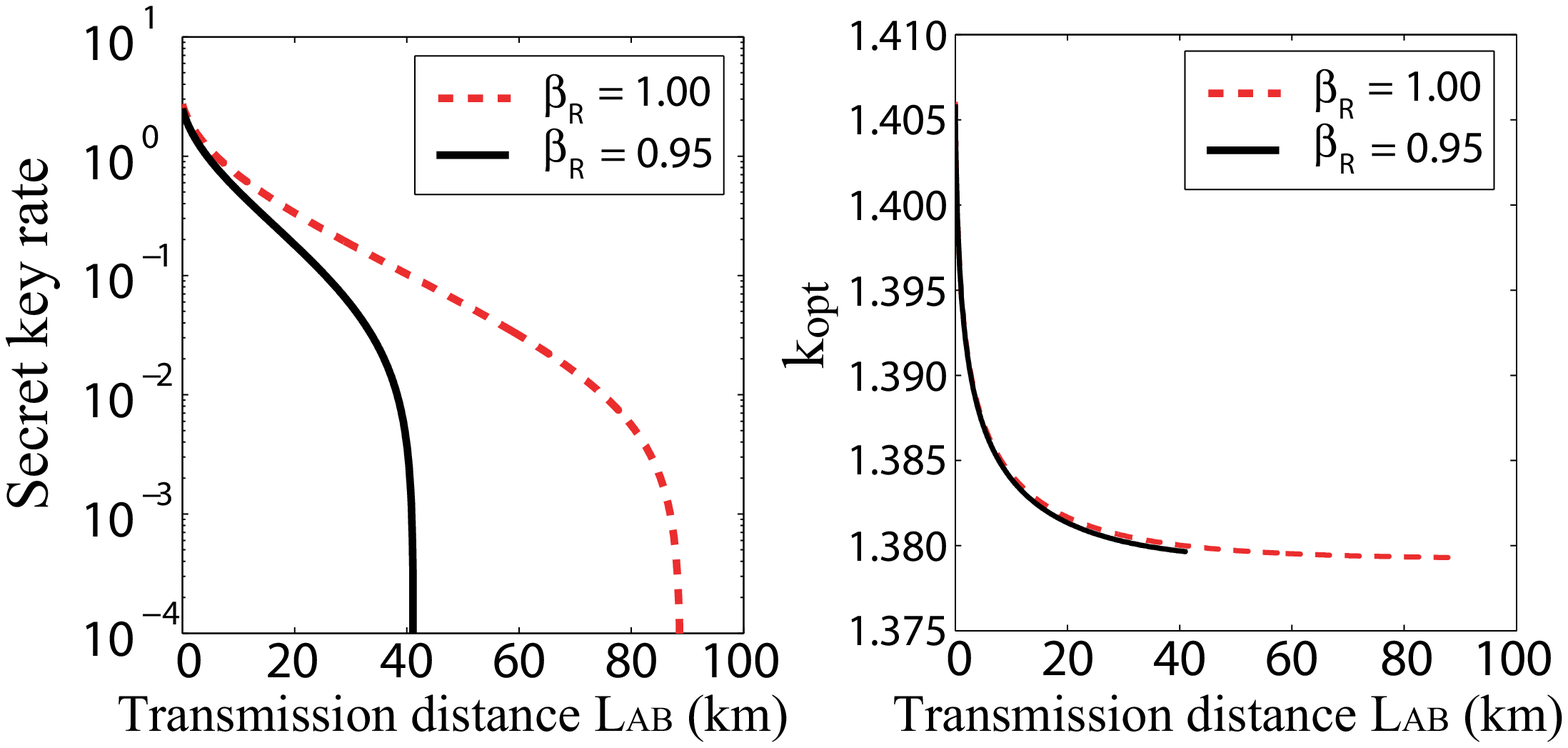}}
\caption{(Color online) (a) PM model of our detection scheme for one-way CV QKD, which contains two steps; the first is the measurement step, and the second is the data-processing step. (b) The left panel is the secret key rate vs transmission distance $L_{AB}$ when using our detection scheme for a one-way protocol with coherent states. The reconciliation efficiency for the red dashed line is $\beta_R = 1$, and for black solid line it is $\beta_R = 0.95$. The right panel is $k_{opt}$ vs $L_{AB}$. The simulation is under the conditions $V_A = V_B = 40$ and $\varepsilon_A = \varepsilon_B = 0.002$ \cite{ZeroTransDis}.}\label{NDect}
\end{figure}

Figure \ref{sim2}(a) shows the secure key rate in an asymmetric case ($L_{AC}\ne L_{BC}$). When $L_{BC}$ increases, the maximal $L_{AC}$ decrease dramatically. If Charlie's position can be close to Bob, the total transmission distance $L_{AB}$ can be a relatively longer distance, up to 80km in theory [the top black solid line in Fig. \ref{sim2}(b)]. Thus, the asymmetric case of CV-MDI QKD indicates the detection scheme that will be discussed below.

The detection scheme discussed here is the same as the EB scheme of CV-MDI QKD except that Bob takes over Charlie's operations in Fig.1(a) with $L_{BC}=0$ km.
Figure \ref{NDect}(a) shows the PM model of this detection scheme, which contains two steps, i.e., the measurement step and the data-processing step. The measurement step is a modified heterodyne detection in which the vacuum state introduced by the BS in the standard heterodyne detection is replaced by a Gaussian-modulated coherent state $\left| {{\alpha _B}} \right\rangle$. The data-processing step uses the measurement results $\{X_C, P_D\}$ and the means $\{x_B, p_B\}$ of the quadratures of $\left| {{\alpha _B}} \right\rangle$ to construct Bob's final data $\{X_B = x_B +kX_C, P_B = p_B-kP_D\}$, where $k$ is the amplification coefficient and will be traversed to find an optimal value $k_{opt}$ which makes the secret key rate the highest. The new detection scheme has two advantages. First, it can defend against all collective attacks against detectors, which is the most important feature. Second, the traversing $k$ can be done classically.

In a practical experiment, the reconciliation efficiency is not 100\%. Figure \ref{NDect}(b) shows the results for a practical reconciliation efficiency $\beta_R = 0.95$ \cite{Jouguet_NPhoton_2013}; the maximal transmission distance using our detection scheme can be 40km.

%Additionally, we considered an almost perfect detection efficiency in our simulation.
%A practical homodyne detector has imperfections such as finite efficiency and electronic noise \cite{PhysRevA.76.042305_ImperfDet}, which will decline the final key rate. For instance, if the detector's efficiency reduces to 90\%, even though the electronic noise is zero the transmission distance will be less than 10km \cite{imperfectD}. However, these imperfections can be compensated by optical preamplifiers \cite{Fossier_JPhysB_2009_PSA, TWOA_2013, Tong_PIA_exp_2012}.

In the above analysis, almost perfect detection efficiency was assumed in our simulation. Next, we will consider a practical homodyne detector which has imperfections such as finite efficiency and electronic noise \cite{PhysRevA.76.042305_ImperfDet}.
These imperfections will increase the ``equivalent excess noise,'' i.e., ${\varepsilon _{{\rm{ImD}}}} = {\varepsilon _A} + {{\left[ {{\eta _B}\left( {{\varepsilon _B} - 2} \right) + 2 + 2{\chi _{Det}}} \right]} \mathord{\left/
 {\vphantom {{\left[ {{\eta _B}\left( {{\varepsilon _B} - 2} \right) + 2 + 2{\chi _{Det}}} \right]} {{\eta _A}}}} \right.
 \kern-\nulldelimiterspace} {{\eta _A}}} = \varepsilon ' + {{2{\chi _{Det}}} \mathord{\left/
 {\vphantom {{2{\chi _{Det}}} {{\eta _A}}}} \right.
 \kern-\nulldelimiterspace} {{\eta _A}}}$ , which holds when traversing the displacement gain \emph{g} to minimize it under two independent entangling
 cloner attacks. Here ${\chi _{Det}} = {{\left( {1 - {\eta _D}} \right)} \mathord{\left/
 {\vphantom {{\left( {1 - {\eta _D}} \right)} {{\eta _D}}}} \right.
 \kern-\nulldelimiterspace} {{\eta _D}}} + {{{\upsilon _{el}}} \mathord{\left/
 {\vphantom {{{\upsilon _{el}}} {{\eta _D}}}} \right.
 \kern-\nulldelimiterspace} {{\eta _D}}}$,  ${\eta _D}$ is the detector's efficiency, and  ${\upsilon _{el}}$ is the variance of the electronic noise. Thus the secret key rate will decrease when the detector is imperfect.
 For instance, if the detector's efficiency reduces to 90\%, even though the electronic noise is zero, the transmission distance will be less than 10 km. Actually, to get a nonzero transmission distance the detector's efficiency has to be above 85.5\% \cite{imperfectD}. However, these imperfections can be compensated for by optical preamplifiers \cite{Fossier_JPhysB_2009_PSA, TWOA_2013, Tong_PIA_exp_2012}.

%Although a practical homodyne detector has imperfections such as finite efficiency and electronic noise \cite{PhysRevA.76.042305_ImperfDet}, these imperfections can be compensated by optical preamplifiers \cite{Fossier_JPhysB_2009_PSA, TWOA_2013, Tong_PIA_exp_2012}.

On the other hand, the detection scheme discussed here is very effective against the local oscillator fluctuation attack \cite{MaXiangchun_Phys.Rev.A_2013_LOfluc} and the calibration attack \cite{Jouguetn_PhysRevA_2013}. The main idea of these two kinds of attacks is that Eve can control the scale of the measurement result by manipulating the LO. If the correct measurement result is denoted by $X_O$, then Eve can force Bob to get a fake measurement result $X_O'=\sqrt{\eta}X_O$, referring to Eq. (3) in \cite{MaXiangchun_Phys.Rev.A_2013_LOfluc}.
When using the detection scheme, the only measurement results are $\{X_C, P_D\}$. When Eve employs those two attacks, Bob will get fake results $\{\sqrt{\eta}X_C, \sqrt{\eta}P_D\}$ and amplify them by the amplification coefficient $k$ in the data-processing step. Because $k$ will be traversed for all possible values to get the highest secret key rate, premultiplying a proportional coefficient $\sqrt{\eta}$ for the measurement results by Eve will only lead to a different optimal value of $k$ while the highest secret key rate is still the same. This can be proven easily because traversing $k$ equals traversing $k\sqrt{\eta}$ when $\eta$ is a constant. Therefore, by using our detection scheme, the final secret key rate under these two kinds of attacks not only is never overestimated but also remains the same as in the nonattack case.

\section{Conclusion}
In this paper, we proposed the continuous-variable measurement-device-independent quantum key distribution protocol, which is immune to all collective attacks against detectors.
The numerical simulation under the entangling cloner attack indicates that the transmission distance between Alice and Bob is limited in the symmetric case ($L_{AC} = L_{BC}$). However, when Charlie is close to Bob, the maximal total transmission distance can still reach 80 km for $\beta_R = 1$ and 40 km for $\beta=0.95$. A corresponding detection scheme for the one-way protocol was proposed which is immune to all collective attacks against detectors. This detection scheme only requires slight revisions to the existing CV-QKD systems and thus shows its feasibility.

\emph{Notes added}: Recently, we become aware of an independent work on the same subject \cite{Pirandola_arXiv.1312.4104_CVMDI}. Those authors use a different security analysis method based on conditional scenarios, which requires a relatively complex postprocessing technique. They also propose a model to describe Eve's general attack and find the most powerful attack, which is very helpful for understanding Eve's attack strategy.

\section{Acknowledgments}
We would like to thank H.-K. Lo, J.-Z. Huang, S. Pirandola, and C. Weedbrook for the helpful discussions. This work is supported by the National Science Fund for Distinguished Young Scholars of China (Grant No. 61225003), the National Natural Science Foundation of China
(Grant No. 61101081), the National Hi-Tech Research and Development (863) Program,
and the Fundamental Research Funds for the Central Universities.

\appendix
\section{Equivalence between PM scheme and EB scheme}

In CV QKD, the generation of a Gaussian-distributed coherent state is usually modeled by measuring one mode of a two-mode squeezed state using heterodyne detection and projecting another mode onto a coherent state. Therefore if we modify the PM scheme by replacing the state preparation step by two other steps, of which the first is Alice and Bob independently generate two TMS states $\hat{\rho}_{A_1A_2}$ and $\hat{\rho}_{B_1B_2}$ and the second is they measure modes $A_1$ and $B_1$ by heterodyne detections, then the modified protocol is equivalent to the original PM scheme. Next, we will show the equivalence between the modified protocol and the EB scheme.

Suppose Alice's initial state is $\hat{\rho}_{A_0A_1A_2}$, where mode $A_0$ is the vacuum state introduced by the heterodyne detection. Bob's initial state is $\hat{\rho}_{B_0B_1B_2}$, where mode $B_0$ is the vacuum state introduced by the heterodyne detection. The initial density matrix of whole system including Eve can be written as
${\hat{\rho} _0} = {\hat{\rho} _{{A_0}{A_1}{A_2}}} \otimes {\hat{\rho} _E} \otimes {\hat{\rho} _{{B_0}{B_1}{B_2}}}$. $\hat{\rho} _E$ may contain multimodes.

For the modified protocol, suppose Alice's initial measurement results are $\{x_A, p_A\}$ and Bob's are $\{x_B', p_B'\}$, then the state Eve gets is
\begin{widetext}
\begin{equation}
\hat{\rho} _{{A_2}{B_2}E}^{\left( {{x_A}{p_A},{x_B'}{p_B'}} \right)} = \frac{{\left\langle {{x_A},{p_A},{x_B'},{p_B'}} \right|U_{{A_1}{A_0}}^{BS}U_{{B_1}{B_0}}^{BS}{\hat{\rho} _0}U_{{A_1}{A_0}}^{BS\dag }U_{{B_1}{B_0}}^{BS\dag }\left| {{x_A},{p_A},{x_B'},{p_B'}} \right\rangle }}{{p\left( {{x_A},{p_A},{x_B'},{p_B'}} \right)}}
\end{equation}
where $p\left( {{x_A},{p_A},{x_B'},{p_B'}} \right)$ ($p_{AB}$ for short) is the probability of getting measurement results $\{{x_A},{p_A},{x_B'},{p_B'}\}$. Eve will get the measurement results $X_C$ and $P_D$ according to a positive operator-valued measurement $\left\{ {\left| {{X_C},{P_D}} \right\rangle \left\langle {{X_C},{P_D}} \right|} \right\}$ on two certain modes, after applying a unitary operation $U_{A_2B_2E}$. Thus the joint probability of $\left\{ {{x_A},{p_A},{x_B'},{p_B'},{X_C},{P_D}} \right\}$ is
\begin{eqnarray}
{p_{ABE}} &=& \Pr \left( {{x_A},{p_A},{x_B'},{p_B'},{X_C},{P_D}} \right)_M
 = {p_{AB}}\left\langle {{X_C},{P_D}} \right|{U_{{A_2}{B_2}E}}\hat{\rho} _{{A_2}{B_2}E}^{\left( {{x_A}{p_A},{x_B'}{p_B'}} \right)}U_{{A_2}{B_2}E}^\dag \left| {{X_C},{P_D}} \right\rangle  \nonumber \\
 &=&  \left\langle {{x_A},{p_A},{x_B'},{p_B'},{X_C},{P_D}} \right|U_{{A_1}{A_0}}^{BS}U_{{B_1}{B_0}}^{BS}{U_{{A_2}{B_2}E}}{\hat{\rho} _0}U_{{A_2}{B_2}E}^\dag U_{{B_1}{B_0}}^{BS\dag }U_{{A_1}{A_0}}^{BS\dag }\left| {{x_A},{p_A},{x_B'},{p_B'},{X_C},{P_D}} \right\rangle
\end{eqnarray}

Because in the last step Bob modifies his data by ${X_B} = {x_B'} + {k_1}{{X}_C},{P_B} = {p_B'} - {k_1}{{P}_D}$, the joint probability of final data $\left\{ {{X_A},{P_A},{X_B},{P_B},{X_C},{P_D}} \right\}$ is
\begin{eqnarray}
{p_{ABE}'} &=& \Pr {\left( {{X_A},{P_A},{X_B},{P_B},{X_C},{P_D}} \right)_D} = \Pr {\left( {{X_A},{P_A} ,{X_B} - {k_1}{X_C},{P_B}+ {k_1}{P_D},{X_C},{P_D}} \right)_M} \nonumber\\
 &=& \left\langle {{X_A},{P_A} ,{X_B} - {k_1}{X_C},{P_B}+ {k_1}{P_D},{X_C},{P_D}} \right|U_{{A_1}{A_0}}^{BS}U_{{B_1}{B_0}}^{BS}{U_{{A_2}{B_2}E}}{\hat{\rho} _0}U_{{A_2}{B_2}E}^\dag U_{{B_1}{B_0}}^{BS\dag }U_{{A_1}{A_0}}^{BS\dag } \nonumber  \\
  &\ &  \ \   \times \left| {{X_A},{P_A},{X_B} - {k_1}{X_C},{P_B} + {k_1}{P_D},{X_C},{P_D}} \right\rangle \nonumber \\
 &=& \left\langle {{X_A},{P_A},{X_B},{P_B},{X_C},{P_D}} \right|{{\hat D}_{{B_1}}}\left( {{k_1}{X_C}} \right){{\hat D}_{{B_0}}}\left( { - i{k_1}{P_D}} \right)U_{{A_1}{A_0}}^{BS}U_{{B_1}{B_0}}^{BS}{U_{{A_2}{B_2}E}}{\hat{\rho} _0}U_{{A_2}{B_2}E}^\dag U_{{B_1}{B_0}}^{BS\dag }U_{{A_1}{A_0}}^{BS\dag } \nonumber \\
 &\ & \ \ \times {\hat D}_{{B_1}}^\dag {\left( {{k_1}{X_C}} \right)}{\hat D}_{{B_0}}^\dag \left( { - i{k_1}{P_D}} \right)\left| {{X_A},{P_A},{X_B},{P_B},{X_C},{P_D}} \right\rangle.
\end{eqnarray}

The measurement applied on mode $B_1$ or $B_0$ is homodyne detection (measuring the \emph{x} quadrature of $B_1$ or the \emph{p} quadrature of $B_0$); thus an extra displacement of the \emph{p} quadrature on $B_1$ or \emph{x} quadrature on $B_0$ will not change the joint probability $p_{ABE}'$. Therefore
\begin{eqnarray}
{p_{ABE}'} &=& \left\langle {{X_A},{P_A},{X_B},{P_B},{X_C},{P_D}} \right|{{\hat D}_{{B_1}}}\left( {{k_1}\left( {{X_C} + i{P_D}} \right)} \right){{\hat D}_{{B_0}}}\left( { - {k_1}\left( {{X_C} + i{P_D}} \right)} \right)U_{{A_1}{A_0}}^{BS}U_{{B_1}{B_0}}^{BS}{U_{{A_2}{B_2}E}}{\hat{\rho} _0} \nonumber \\
&\ & \ \ \times U_{{A_2}{B_2}E}^\dag U_{{A_1}{A_0}}^{BS\dag }U_{{B_1}{B_0}}^{BS\dag }{\hat D}_{{B_1}}^\dag {\left( {{k_1}\left( {{X_C} + i{P_D}} \right)} \right)}{\hat D}_{{B_0}}^\dag \left( { - {k_1}\left( {{X_C} + i{P_D}} \right)} \right)\left| {{X_A},{P_A},{X_B},{P_B},{X_C},{P_D}} \right\rangle
\end{eqnarray}

Suppose the quadratures of mode $B_0$ and $B_1$ are $\hat{x}_{B_0}, \hat{p}_{B_0}$ and $\hat{x}_{B_1}, \hat{p}_{B_1}$. After passing through the 50:50 BS first and then two displacements ${{\hat D}_{{B_1}}}\left( {{k_1}\left( {{X_C} + i{P_D}} \right)} \right), {{\hat D}_{{B_0}}}\left( { - {k_1}\left( {{X_C} + i{P_D}} \right)} \right)$, these quadratures become $\hat{x}_{A_0}', \hat{p}_{A_0}'$ and $\hat{x}_{A_1}', \hat{p}_{A_1}'$. Then
\begin{equation}
\left\{ \begin{array}{l}
{{\hat x}_{{B_0}}}' = \frac{1}{{\sqrt 2 }}\left( {{{\hat x}_{{B_0}}} + {{\hat x}_{{B_1}}}} \right) + {k_1}{X_C} = \frac{1}{{\sqrt 2 }}\left( {{{\hat x}_{{B_0}}} + \left( {{{\hat x}_{{B_1}}} + \sqrt 2 {k_1}{X_C}} \right)} \right)\\
{{\hat p}_{{B_0}}}' = \frac{1}{{\sqrt 2 }}\left( {{{\hat p}_{{B_0}}} + {{\hat p}_{{B_1}}}} \right) + {k_1}{P_D} = \frac{1}{{\sqrt 2 }}\left( {{{\hat p}_{{B_0}}} + \left( {{{\hat p}_{{B_1}}} + \sqrt 2 {k_1}{P_D}} \right)} \right)\\
{{\hat x}_{{B_1}}}' = \frac{1}{{\sqrt 2 }}\left( {{{\hat x}_{{B_0}}} - {{\hat x}_{{B_1}}}} \right) - {k_1}{X_C} = \frac{1}{{\sqrt 2 }}\left( {{{\hat x}_{{B_0}}} - \left( {{{\hat x}_{{B_1}}} + \sqrt 2 {k_1}{X_C}} \right)} \right)\\
{{\hat p}_{{B_1}}}' = \frac{1}{{\sqrt 2 }}\left( {{{\hat p}_{{B_0}}} - {{\hat p}_{{B_1}}}} \right) - {k_1}{P_D} = \frac{1}{{\sqrt 2 }}\left( {{{\hat p}_{{B_0}}} - \left( {{{\hat p}_{{B_1}}} + \sqrt 2 {k_1}{P_D}} \right)} \right)
\end{array} \right.
\end{equation}
They are the same as firstly displacing ${{\hat D}_{{B_1}}}\left( \sqrt{2}{{k_1}\left( {{X_C} + i{P_D}} \right)} \right)$ on mode $B_1$ and then passing through the BS. Thus, we have
\begin{eqnarray}
{p_{ABE}'} &=& \left\langle {{X_A},{P_A},{X_B},{P_B},{X_C},{P_D}} \right|\left[ {U_{{B_1}{B_0}}^{BS}{{\hat D}_{{B_1}}}\left( {\sqrt 2 {k_1}\left( {{X_C} + i{P_D}} \right)} \right)} \right]U_{{A_1}{A_0}}^{BS}{U_{{A_2}{B_2}E}}{\hat{\rho} _0} \nonumber \\
 &\ & \ \ \times U_{{A_2}{B_2}E}^\dag U_{{A_1}{A_0}}^{BS\dag }\left[ {{\hat D}_{{B_1}}^\dag \left( {\sqrt 2 {k_1}\left( {{X_C} + i{P_D}} \right)} \right)U_{{B_1}{B_0}}^{BS\dag }} \right]\left| {{X_A},{P_A},{X_B},{P_B},{X_C},{P_D}} \right\rangle. \label{PMprob}
\end{eqnarray}

For the EB scheme, Eve does the measurement first. We make the same assumption as above; then after Eve gets the measurement results $\{X_C, P_D\}$, the state left for Alice and Bob will be
\begin{eqnarray}
\hat \rho _{{A_0}{A_1}{B_0}{B_1}}^{\left( {{X_C},{P_D}} \right)} = \frac{{\left\langle {{X_C},{P_D}} \right|{U_{{A_2}{B_2}E}}{\hat{\rho} _0}U_{{A_2}{B_2}E}^\dag \left| {{X_C},{P_D}} \right\rangle }}{{p\left( {{X_C},{P_D}} \right)}},
\end{eqnarray}
where $p\left( {{X_C},{P_D}} \right)$ is the probability of getting measurement results $\{X_C, P_D\}$.
Then Bob will displace mode $B_1$ first by ${{{\hat D}_{{B_1}}}\left( {g\left( {{X_C} + i{P_D}} \right)} \right)}$ and measure it using a heterodyne detector. The probability of getting the final data $\{X_A, P_A, X_B, P_B\}$ given $\{X_C,P_D\}$ is
\begin{eqnarray}
{p_{AB|CD}} &=& \Pr \left( {{X_A},{P_A},{X_B},{P_B}|{X_C},{P_D}} \right) \nonumber = \left\langle {{X_A},{P_A},{X_B},{P_B}} \right|U_{{B_1}{B_0}}^{BS}{{\hat D}_{{B_1}}}\left( {g\left( {{X_C} + i{P_D}} \right)} \right)U_{{A_1}{A_0}}^{BS} \nonumber \\
&\ & \ \ \times \hat \rho _{{A_0}{A_1}{B_0}{B_1}}^{\left( {{X_C},{P_D}} \right)}U_{{A_1}{A_0}}^{BS\dag }{\hat D}_{{B_1}}^\dag \left( {g\left( {{X_C} + i{P_D}} \right)} \right)U_{{B_1}{B_0}}^{BS\dag }\left| {{X_A},{P_A},{X_B},{P_B}} \right\rangle.
\end{eqnarray}
The joint probability of all variables $\left\{ {{X_A},{P_A},{X_B},{P_B},{X_C},{P_D}} \right\}$ is
\begin{eqnarray}
{p_{ABE}''} &=& {p_{AB|CD}}p\left( {{X_C},{P_D}} \right) = \left\langle {{X_A},{P_A},{X_B},{P_B},{X_C},{P_D}} \right|\left[ {U_{{B_1}{B_0}}^{BS}{{\hat D}_{{B_1}}}\left( {g\left( {{X_C} + i{P_D}} \right)} \right)} \right]U_{{A_1}{A_0}}^{BS} \nonumber \\
&\ & \ \ \times {U_{{A_2}{B_2}E}}{{\hat \rho }_0}U_{{A_2}{B_2}E}^\dag U_{{A_1}{A_0}}^{BS\dag }\left[ {{\hat D}_{{B_1}}^\dag \left( {g\left( {{X_C} + i{P_D}} \right)} \right)U_{{B_1}{B_0}}^{BS\dag }} \right]\left| {{X_A},{P_A},{X_B},{P_B},{X_C},{P_D}} \right\rangle. \label{EBprob}
\end{eqnarray}

It is easy to check that the two joint probabilities ${p_{ABE}'}$ and ${p_{ABE}''}$ are the same once we let $g$ equal $\sqrt{2} k_1$. Therefore the EB scheme is equal to the modified protocol, and because of the equivalence between the modified protocol and the PM scheme, the EB scheme further equals the PM scheme.

In the modified protocol, if Bob's initial measurement results are $\{x_B', p_B'\}$, then the coherent state sent out from him is $\left| {{x_A} + i{p_A}} \right\rangle$, where $x_A = \sqrt{2\frac{V_B - 1}{V_B+1}}x_A'$ and $p_A = \sqrt{2\frac{V_B - 1}{V_B+1}}p_A'$. So the amplification coefficient $k$ in the PM scheme is described as $k = \frac{k_1}{\sqrt{2\frac{V_B - 1}{V_B+1}}} = \frac{g}{\sqrt{\frac{V_B - 1}{V_B+1}}}$.

\section{Relationship of quadratures used in numerical simulation}
After passing through the channels,
\begin{equation}
{\left\{ {\begin{array}{*{20}{l}}
{\hat A' = \sqrt {{\eta _A}} {{\hat A}_2} + \sqrt {1 - {\eta _A}} {{\hat E}_2}}\\
{\hat B' = \sqrt {{\eta _B}} {{\hat B}_2} + \sqrt {1 - {\eta _B}} {{\hat E}_5}}
\end{array}} \right.}.
\end{equation}

Modes $A'$ and $B'$ interfere on the 50:50 BS, then modes $C$ and $D$ are
\begin{equation}
{\left\{ {\begin{array}{*{20}{l}}
{\hat C = \frac{1}{{\sqrt 2 }}\left( { \hat A' - \hat B'} \right) = \frac{1}{{\sqrt 2 }}\left( { \sqrt {{\eta _A}} {{\hat A}_2} - \sqrt {{\eta _B}} {{\hat B}_2}} \right) + \frac{1}{{\sqrt 2 }}\left( {  \sqrt {1 - {\eta _A}} {{\hat E}_2} - \sqrt {1 - {\eta _B}} {{\hat E}_5}} \right)}\\
{\hat D = \frac{1}{{\sqrt 2 }}\left( {\hat A' + \hat B'} \right) = \frac{1}{{\sqrt 2 }}\left( {\sqrt {{\eta _A}} {{\hat A}_2} + \sqrt {{\eta _B}} {{\hat B}_2}} \right) + \frac{1}{{\sqrt 2 }}\left( {\sqrt {1 - {\eta _A}} {{\hat E}_2} + \sqrt {1 - {\eta _B}} {{\hat E}_5}} \right)}
\end{array}} \right.}.
\end{equation}

After the displacement, mode $B_1'$ becomes
\begin{equation}
{\left\{ {\begin{array}{*{20}{l}}
{{{\hat B}_{1x}}^\prime  = {{\hat B}_{1x}} + g{{\hat C}_x} = \left( {{{\hat B}_{1x}} - g\frac{{\sqrt {{\eta _B}} }}{{\sqrt 2 }}{{\hat B}_{2x}}} \right) + g\frac{{\sqrt {{\eta _A}} }}{{\sqrt 2 }}{{\hat A}_{2x}} + \frac{g}{{\sqrt 2 }}\left( {  \sqrt {1 - {\eta _A}} {{\hat E}_{2x}} - \sqrt {1 - {\eta _B}} {{\hat E}_{5x}}} \right)}\\
{{{\hat B}_{1p}}^\prime  = {{\hat B}_{1p}} + g{{\hat D}_p} = \left( {{{\hat B}_{1p}} + g\frac{{\sqrt {{\eta _B}} }}{{\sqrt 2 }}{{\hat B}_{2p}}} \right) + g\frac{{\sqrt {{\eta _A}} }}{{\sqrt 2 }}{{\hat A}_{2p}} + \frac{g}{{\sqrt 2 }}\left( {\sqrt {1 - {\eta _A}} {{\hat E}_{2p}} + \sqrt {1 - {\eta _B}} {{\hat E}_{5p}}} \right)}
\end{array}} \right.}.
\end{equation}
\end{widetext}

%\section{Calculation of secret key rate $K_2^R$}
%Suppose the covariance matrix $\gamma_{A_1B_1'}$ is
%\begin{equation}
%\left( {\begin{array}{*{20}{c}}
% {{V_A}\cdot{{\rm{I}}_2}}&{\sqrt {T\left( {V_A^2 - 1} \right)} \cdot{\sigma _z}}\\
%{\sqrt {T\left( {V_A^2 - 1} \right)} \cdot{\sigma _z}}&{\left[ {T\left( {{V_A} } + \chi \right) } \right]\cdot{{\rm{I}}_2}}
%\end{array}} \right).
%\end{equation}
%
%The classical mutual information $I\left( {{X_A, P_A}:{X_B, P_B}} \right)$ is
%\begin{equation}
% \emph{log} \left[ \frac{V_B+1}{V_{B|A}+1}\right] = \emph{log} \left[ \frac{T\left(V+\chi\right)+1}{T\left(\chi + 1\right)+1}\right].
%\end{equation}

\newpage %Just because of unusual number of tables stacked at end

%\bibliography{reference}% Produces the bibliography via BibTeX.
\bibliography{ab}% Produces the bibliography via BibTeX.

\begin{thebibliography}{51}
\bibitem[{Zer()}]{ZeroTransDis}
\bibinfo{note}{`0km' just represents the transmission distance from Bob to
  Charlie is very short but not exact 0 meters. In fact even when Charlie is in
  Bob's side, there still exists a distance between Bob's laser and the
  beamsplitter. Therefore, in the numerical simulation although we assume the
  channel transmittance is $\eta_B = 1$, the excess noise $\varepsilon_B$ still
  exists, and $\varepsilon_B = 0.002$.}
\bibitem[{imp()}]{imperfectD}
\bibinfo{note}{This simulation is based on the asymmetric case, where $L_{BC}$
  is 0km. The variances of Alice's and Bob's EPR and channel parameters are the
  same as Fig. 5. The electronic noise is set to be zero and reconciliation
  efficiency is 1.}
\end{thebibliography}
\end{document}